\documentclass[aps,prl,footinbib,floats,twocolumn,showpacs,superscriptaddress,longbibliography]{revtex4-1}

\usepackage{times}
\usepackage{amsmath}
\usepackage{amsfonts}
\usepackage{amssymb}
\usepackage{graphicx}
\usepackage[gen]{eurosym}
\usepackage{threeparttable}
\usepackage{color}
\usepackage{url}
\newcommand{\nin}[1]{N^{\mbox{\scriptsize in}}_{#1}}
\newcommand{\nout}[1]{N^{\mbox{\scriptsize out}}_{#1}}
\newcommand{\expM}[1]{{\mbox{\scriptsize #1}}}
\renewcommand{\(}{\left(}
\renewcommand{\)}{\right)}

\begin{document}

\title{A model to identify urban traffic congestion hotspots in complex networks}

\author{Albert Sol\'e-Ribalta}
\affiliation{Departament d'Enginyeria Inform\`atica i Matem\`atiques,
Universitat Rovira i Virgili, 43007 Tarragona, Spain}

\author{Sergio G\'omez}
\affiliation{Departament d'Enginyeria Inform\`atica i Matem\`atiques,
Universitat Rovira i Virgili, 43007 Tarragona, Spain}

\author{Alex Arenas}
\affiliation{Departament d'Enginyeria Inform\`atica i Matem\`atiques,
Universitat Rovira i Virgili, 43007 Tarragona, Spain}

\begin{abstract}
Traffic congestion is one of the most notable problems arising in worldwide urban areas, importantly compromising human mobility and air quality. Current technologies to sense real-time data about cities, and its open distribution for analysis, allow the advent of new approaches for improvement and control. Here, we propose an idealized model, the Microscopic Congestion Model, based on the critical phenomena arising in complex networks, that allows to analytically predict congestion hotspots in urban environments. Results on real cities' road networks, considering, in some experiments, real-traffic data, show that the proposed model is capable of identifying susceptible junctions that might become hotspots if mobility demand increases.
\end{abstract}

\maketitle

\section{Introduction}
Urban life is characterized by a huge mobility, mainly motorized. Amidst the complex urban management problems there is a prevalent one: traffic congestion. Several approaches exist to efficiently design road networks\cite{Yang1998} and routing strategies\cite{Bast2007}, however, the establishment of collective actions, given the complex behavior of drivers, to prevent or ameliorate urban traffic congestion is still at its dawn. Usually, congestion is not homogeneously distributed around all city area but there are salient locations where congestion is settled. We call this locations congestion hotspots. These hotspots usually correspond to junctions and are problematic for the efficiency of the network as well as for the health of pedestrians and drivers. It has been shown\cite{Petersson1978ExposureToTrafficExhaust} that drivers in-queue in a traffic jam are the most affected individuals to car exhaust pollution inhalation. In addition, these hotspots are usually located in the city center, magnifying the problem\cite{Raducan2009PolutansBucarest}. Assuming that congestion is an inevitable consequence of urban motorized areas, the challenge is to develop strategies towards a sustainable congestion regime at which delays and pollution are under control. The first step to confront congestion is the modelling and understating of the congestion phenomena.

The modeling of traffic flows it is prevalent hot topic since the late 70's when the Gipps' model appear \cite{wilson2001analysis}. The Gipps' model and other car-following models \cite{treiber2000congested, newell2002simplified} have evidenced the necessity of modeling traffic flows to improve road network efficiency and also have shown how congestion severely affects the traffic flows.  Since ten years ago also the complex networks' community has proposed stylized models to analyze the problem of traffic congestion in networks and design optimal topologies to avoid it\cite{guimera2002optimal, tadic04, donetti05, liang05, ashton05, singh05, danila06, bart06, doro08, kim09, li10, ramasco10, sce10, geeson12}. The focus of attention of the previous works was the onset of congestion, which corresponds to a critical point in a phase transition, and how it depends on the topology of the network and the routing strategies used. However, the proper analysis of the system after congestion has remained analytically slippery. It is known that when a transportation network reaches congestion, the system becomes highly non-linear, large fluctuation exists and the travel time and the amount of vehicles queued in a junction diverge\cite{doro08}. This phenomenon is equivalent to a phase transition in physics, and its modeling is challenging\cite{Bianconi2009Cong, Echenique05EPL, Stanley2015LinkPredictability}. Here, we propose an idealized model to predict the behaviour of transportation networks after the onset of congestion. The presented model is analytically tractable and can be iteratively solved up to convergence. To the best of our knowledge, this is the first analytical model that is able to give predictions beyond the onset of congestion. We present the model in terms of road transportation networks but it could also be applied to analyze other types of transportation networks such as: computer networks, business organizations or social networks.

\section{Results}

\begin{figure*}
	\begin{center}
		\includegraphics[width=0.95\textwidth]{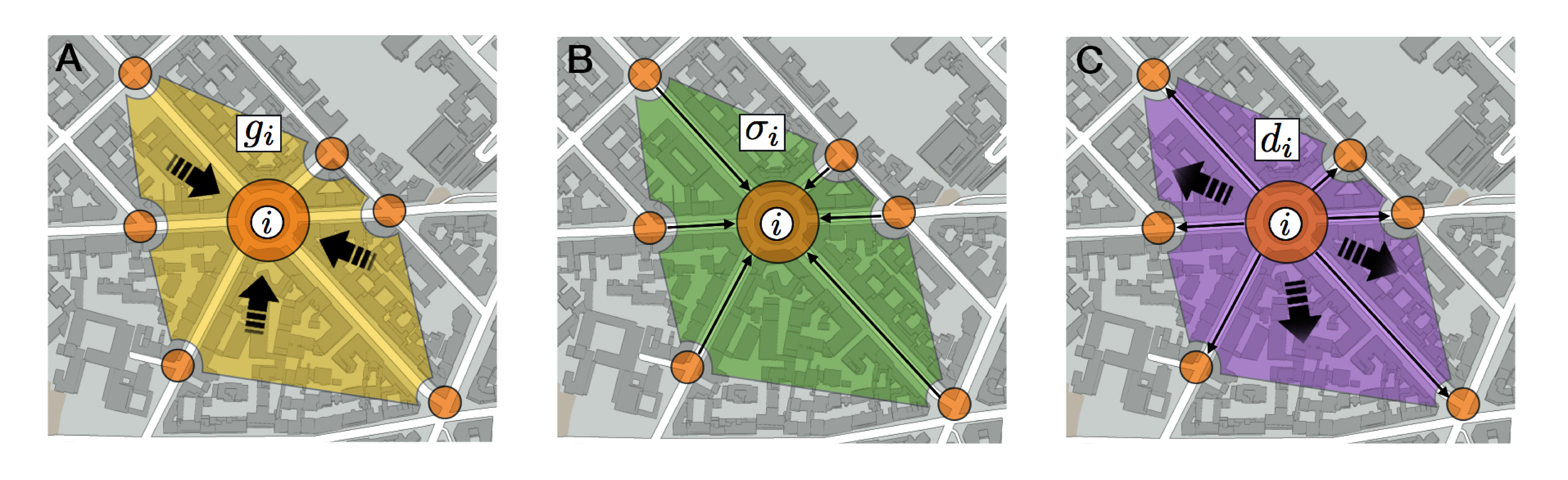}
	\end{center}
        \caption{Illustration of the variables of the MCM model. {\bf (A)} Vehicles entering junction $i$ from the area surrounding $i$. {\bf (B)}  Vehicles entering junction $i$ from its neighboring junctions. {\bf (C)} Vehicles leaving junction $i$, either to go to other neighboring junctions or to finishing the trip in its surrounding area.}
        \label{fig:modelExplanation}
\end{figure*}

To identify congestion hotspots in urban environments we propose a model based on the theory of critical (congestion) phenomena on complex networks. The model, that we call {\em Microscopic Congestion Model} (MCM), is a mechanistic model (yet simple) and analytically tractable. It is based on assuming that the growth of vehicles observed in each congested node of the networks is constant. This usually happens in real transportation networks at the stationary state. The assumption allows us to describe, with a set of balance equations (one for each node), the increment of vehicles in the junction queues' and the number of vehicles arriving or traversing each junction from neighboring junctions. Mathematically, the increment of the vehicles per unit time at every junction $i$ of the city, $\Delta{q}_{i}$, satisfies the following balance equation:
\begin{equation}
	\Delta{q}_{i} = g_{i} + \sigma_{i} - d_{i},
	\label{BE}
\end{equation}
where $g_{i}$ is the average number of vehicles entering junction $i$ from the area surrounding $i$, $\sigma_{i}$ is the average number of vehicles that arrive to junction $i$ from the adjacent links of that junction, and $d_{i} \in [0,\tau_i]$ corresponds to the average of vehicles that actually finish in junction $i$ or traverse towards other junctions. Note that the value of $d_i$ is upper-bounded by the maximum amount of vehicles $\tau_i$ that can traverse junction $i$ in a time step. This simulates the physical constraints of the road network. A graphical explanation of the variables of the model is shown in Fig.~\ref{fig:modelExplanation}.

The system of Eqs.~\ref{BE} defined for every node $i$, is coupled through the incoming flux variables $\sigma_{i}$, that can be expressed as
\begin{equation}
	\sigma_{i} = \sum^{S}_{j=1} P_{ji} p_{j} d_{j},
	\label{sigma}
\end{equation}
where $P_{ji}$ accounts for the routing strategy of the vehicles (probability of going from $j$ to $i$), $p_j$ stands for the probability of traversing junction $j$ but not finishing at $j$ and $S$ is the number of nodes in the network (see {\em Materials and Methods} for a detailed description of the MCM).

For each junction $i$, the onset of congestion is determined by $d_i=\tau_i$, meaning that the junction is behaving at its maximum capability of processing vehicles. Thus, for any flux generation ($g_i$), routing strategy ($P_{ij}$) and origin-destination probability distribution, Eqs.~\ref{BE} can be solved using an iterative approach (see {\em Materials and Methods}) to predict the increase of vehicles per unit time at each junction of the network ($\Delta{q}_{i}$). The only hypothesis we use is that the system dynamics has reached a stationary state in which the growth of the queues is constant. It is worth commenting here that the MCM model considers a fixed average of new vehicles entering the system $g_i$. However, $g_i$ certainly changes during day time, with increasing values in rush hours and lower values during off-peak periods. MCM can easily consider evolving values of $g_i$ provided the time scale to reach the stationary state in the MCM (which is usually of the order of minutes in real traffic systems) is shorter than the rate of change in the evolution of $g_i$ (which is usually of the order of hours for the daily peaks).

\subsection{Validation on synthetic networks}

To validate MCM we conducted experiments on several synthetic networks and with two different routing strategies: local search strategy and shortest path strategy. In both routing strategies we assume, for simplicity, that all vehicles randomly choose the starting and ending junctions of their journey uniformly within all junctions of the network. Thus, each junction generates new vehicles with the same rate $g_i = \rho$. For shortest path strategy, vehicles follow a randomly selected shortest path towards the destination. Without loss of generality we fix $\tau=1$ and analyze the performance of MCM for different values of $\rho$.

\begin{figure}[!ht]
  \begin{center}
	 \begin{tabular}{l}
		\hskip 1em{\bf A} \hskip 16em {\bf B}\\
		\includegraphics[width=0.9\columnwidth]{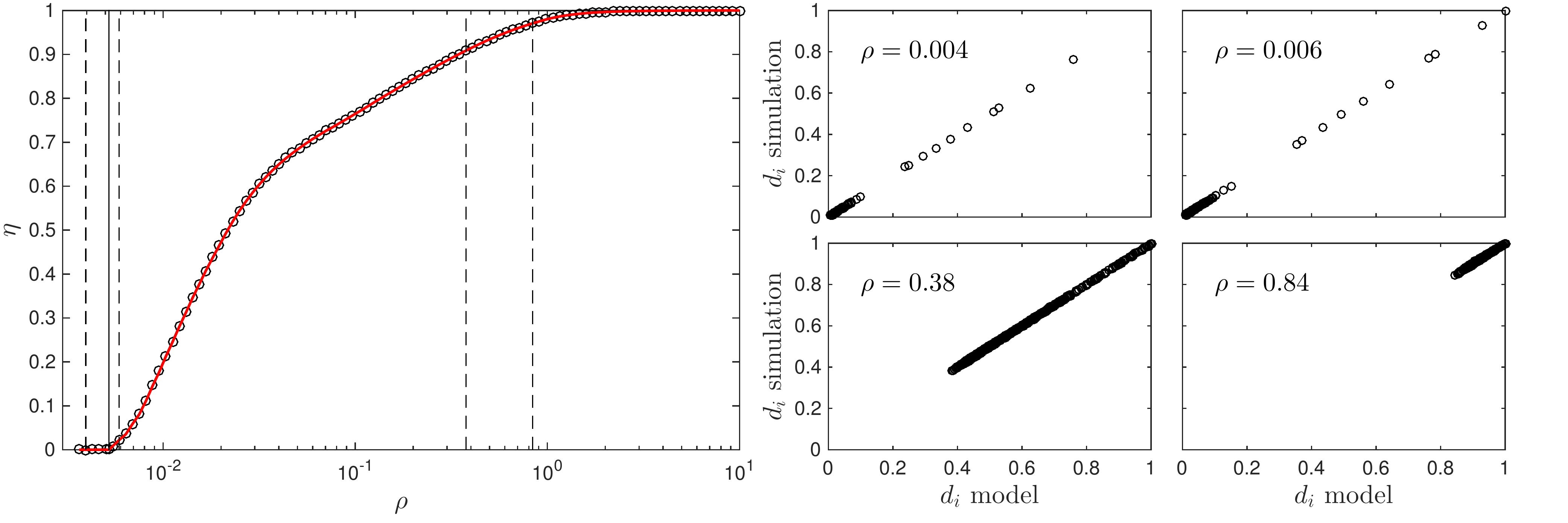}\\
		\hskip 1em{\bf C} \hskip 16em {\bf D}\\
		\includegraphics[width=0.9\columnwidth]{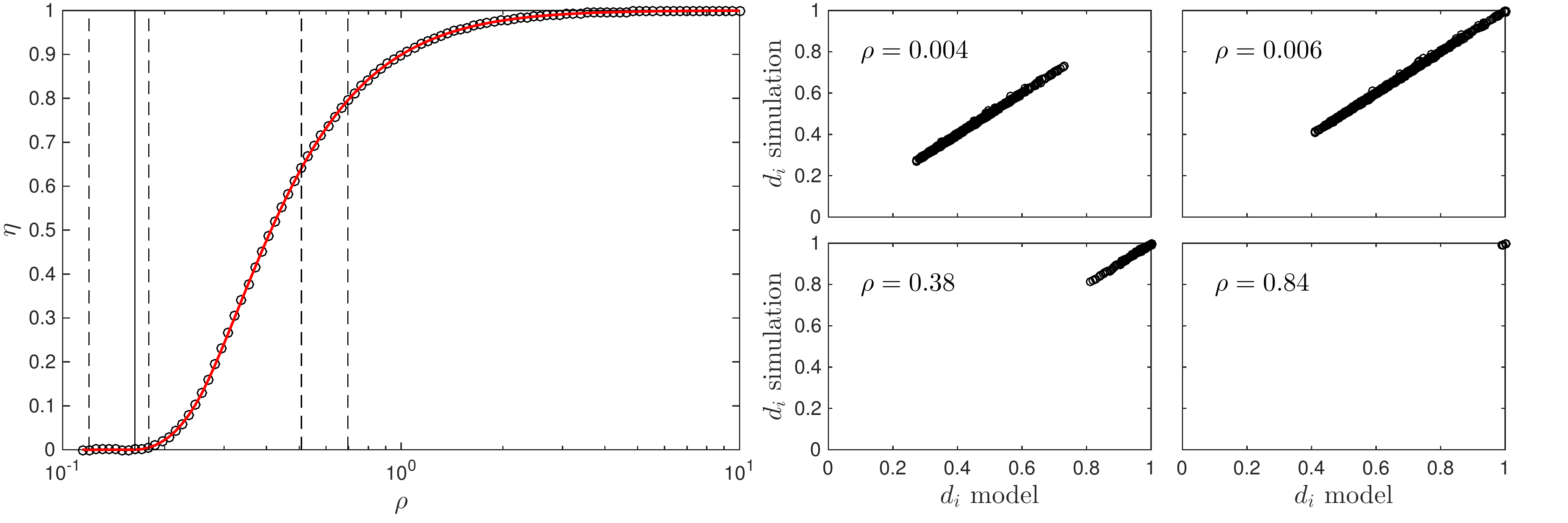}
 	 \end{tabular}
	\end{center}
	\caption{Validation of the Microscopic Congestion Model with Barab{\'a}si-Albert ({\bf A, B}) and Erd{\H{o}}s-R\'{e}nyi ({\bf C, D}) networks of 1000 nodes and shortest path routing strategy. In the construction procedure of the Barab{\'a}si-Albert networks each new node is connected to 1 existing node in the network. The Erd{\H{o}}s-R\'{e}nyi networks have an average degree of 50. In {\bf A} and {\bf C}, accuracy in predicting the order parameter $\eta$. In {\bf B} and {\bf D}, correlation between predicted and simulated values of $d_i$. Vertical solid lines on plot {\bf A} and {\bf C} show the predicted critical generation rate $\rho_c$ (see {\em Materials and Methods}). Vertical dashed lines show the $\rho$ values where $d$ is evaluated on the panels {\bf B} and {\bf D}.}
	\label{fig:validation_sp_N1000}
\end{figure}

Figure~\ref{fig:validation_sp_N1000} shows the accuracy on predicting the values of the order parameter $\eta=\frac{\sum \Delta q_i}{\rho S}$ and $d_i$ for shortest paths routing strategy. As in refs.~\cite{Arenas2010OptimalInfoTransm, Guimera2002OptimalTopologies}, this order parameter $\eta$ corresponds to the ratio between in-transit and generated vehicles. All experiments show that the MCM achieves high accuracy in predicting the macroscopic and microscopic variables of the stylised transportation dynamics.

\subsection{Application to real scenarios}

INRIX Traffic Scorecard (http://www.inrix.com/) reports the rankings of the most congested countries worldwide in 2014. US, Canada and most of the European countries are in the top 15, with averages that range from 14 to 50 hours per year wasted in congestion, with their corresponding economical and environmental negative consequences. To demonstrate that the MCM model can be applied to real scenarios to obtain real predictions, in the following we apply the MCM model to the ninth most congested cities according to the INRIX Traffic Scorecard (see Table~\ref{tab:INRIX_vs_HOTSPOT}).

We first focus on the city of Milan, the city with largest INRIX value. To evaluate the outcome of the MCM model, we first gather data about the road network topology using Open Street Map (OSM). OSM data represents each road (or way) with an ordered list of nodes which can either be road junctions or simply changes of the direction of the road. We have obtained the required abstraction of the road network building a simplified version of the OSM data which only accounts for road junctions (nodes). Then, for each pair of adjacent junctions we have queried the real travel distance (i.e.\ following the road path) using the API provided by Google Maps. The resulting network corresponds to a spatial weighed directed network\cite{barth2011} where the driving directions are represented and the weight of each link indicates the expected traveling time between two adjacent junctions.

Second, we build up the dynamics of the model analyzing real traffic data provided by Telecom Italia for their Big Data Challenge. The data provides, for every car entering the cordon pricing zone in Milan during November and December 2013, an encoding of the car's plate number, time and gate of entrance (a total of  9183475 records). This allows us to obtain the (hourly) average incoming and outgoing traffic flow, for each gate of the cordon taxed area.

\begin{figure*}[!ht]
	 \begin{tabular}{ll}
	 	{\bf A} & {\bf B}\\
		\includegraphics[width=0.90\columnwidth]{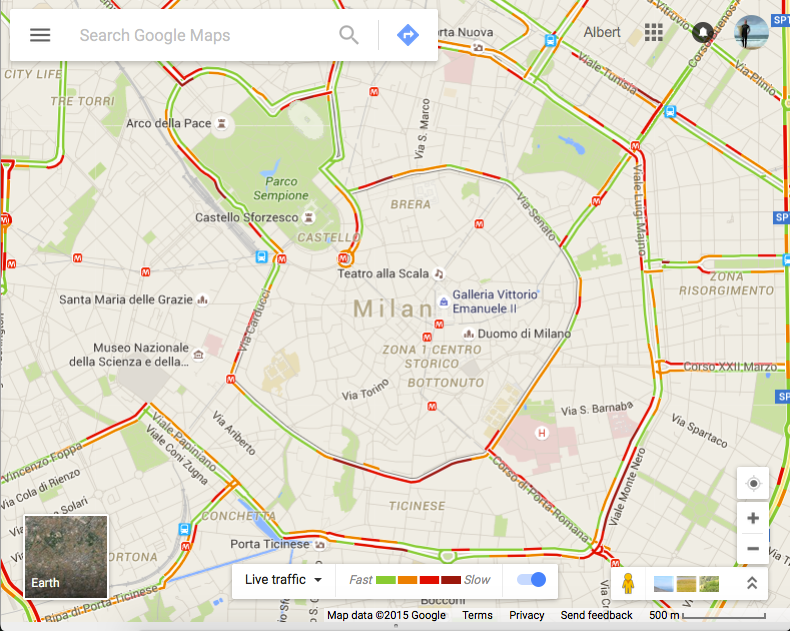} ~~~ &~~~
		\includegraphics[width=0.90\columnwidth]{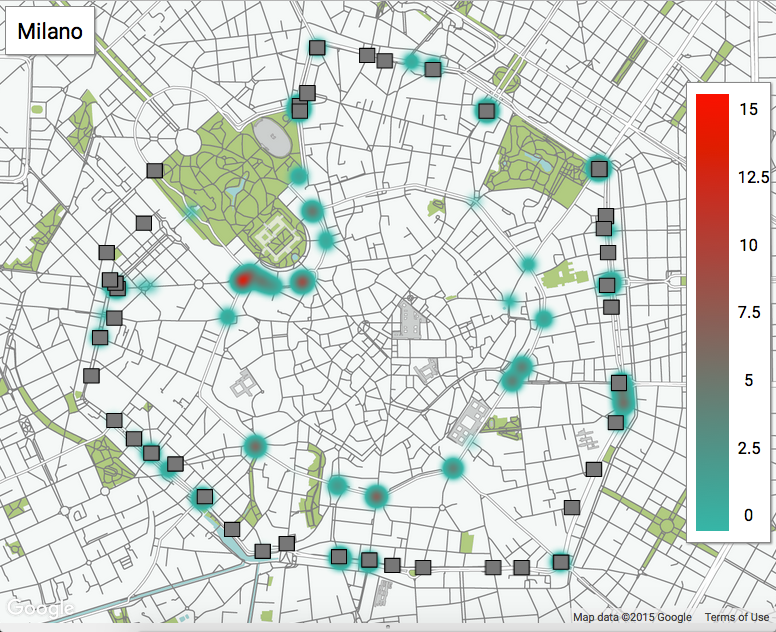}
 	 \end{tabular}
	\caption{Congestion hotspot analysis of the city of Milan. Panel {\bf A} shows the typical situation around 9 a.m. for a week day. The image and the data has been obtained with Google Maps. Google maps displays traffic information considering historical data and real-time car velocity reported by smartphones \cite{barth2009googlemapstraffic}. Panel {\bf B} shows the prediction of the MCM model considering the real road topology obtained using Open Street Map and real traffic data provided by Telecom Italia for their Big Data Challenge. For all congestion hotspots the model has predicted, we shown its mean increment of the queue size, $\langle \Delta q_{i} \rangle$.}
	\label{fig:milano_hotspot_map}
\end{figure*}

\begin{figure}
	\begin{center}
	 \begin{tabular}{ll}
	 	{\bf A} & {\bf B}\\
		\includegraphics[width=0.45\columnwidth]{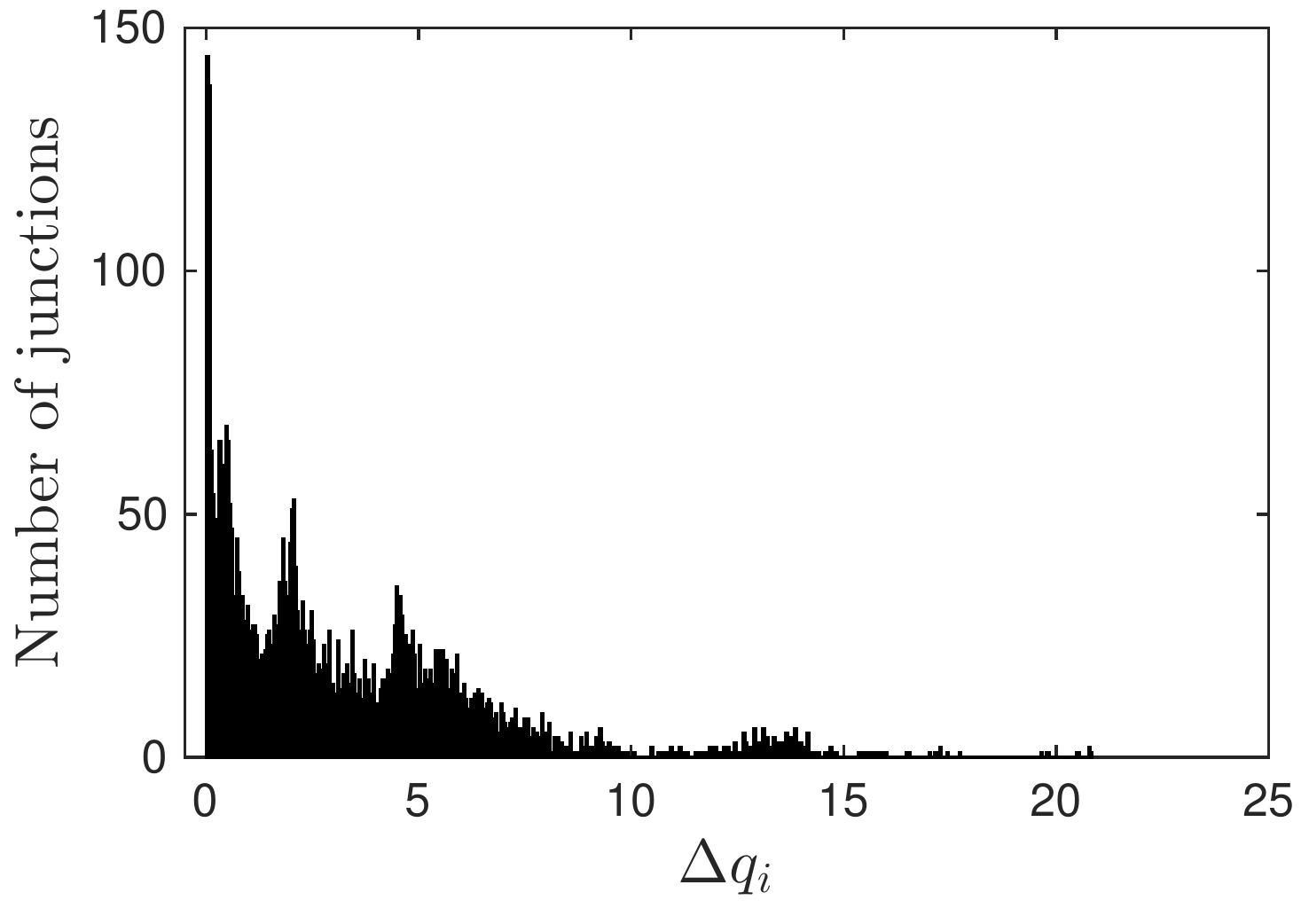} & \includegraphics[width=0.45\columnwidth]{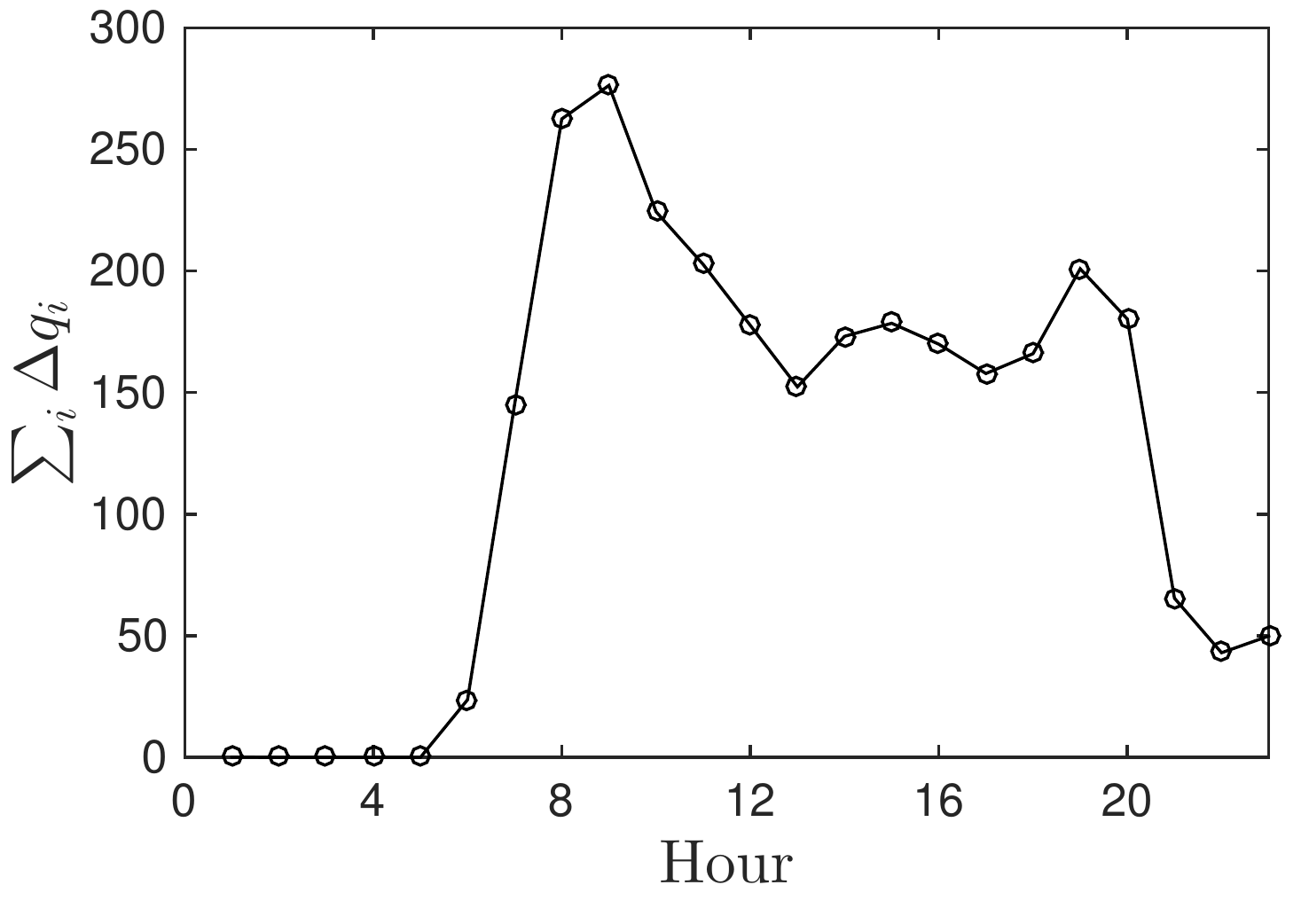}
 	 \end{tabular}
	\end{center}
        \caption{Statistics of Milan congestion hotspots. Panel {\bf A} shows the distribution of the vehicle increments ($\Delta q_i$) of each congestion hotspot predicted by the MCM. The plot aggregates the MCM predicted congestion considering the average traffic for every week day and every hour of the day.  Panel {\bf B} shows the average city congestion in terms of mean increment of vehicle in transit for every hour of the day during a week day.}
        \label{fig:milan_statistics}
\end{figure}

Given the previous topology and traffic information, we generated traffic compatible with the observations, and evaluated the outcome of the MCM model. Specifically, the simulated dynamics is as follows: for each vehicle entering the Area-C we fix a randomly selected location as destination and use the shortest path route towards it. After the vehicle has arrived to its destination, it randomly chooses an exit door and travels to it also using the shortest path route. This is similar to the well-known \textit{Home-to-Work} travel pattern (see details in {\em Materials and Methods}). Figures~\ref{fig:milano_hotspot_map} and~\ref{fig:milan_statistics} show the obtained results. Figures~\ref{fig:milano_hotspot_map}{\bf B} displays the predicted congestion hotspots on a map of Milan, panel {\bf A} of the same figure shows a real traffic situation obtained with google maps. We see that the predicted congestion hotspots are located in the circular roads of Milan as well as on the arterial roads of the city; this agrees with the real traffic situation shown in panel {\bf A}.  Figures~\ref{fig:milan_statistics}{\bf A} shows the distribution of the mean increments each junction has to deal with. This might be a good indicator to decide about future planning actuations to improve city mobility. However, differently from what is described in \cite{liang05}, the improvement of the throughput of a single junction might not be enough to improve city mobility since this might end up with the collapse of neighbouring junctions (their incoming rate $\sigma_i$ will increase). This is situation is similar to the Braess' paradox \cite{youn2008price}. Figures~\ref{fig:milan_statistics}{\bf B} shows the mean increment of vehicles (in vehicles per minute) for each hour of the weekday. The figure clearly shows the morning and evening rush hours as well as the lunch time.

For the other top nine congested cities, we do not have previous traffic information, neither about the real flux of vehicles nor about the vehicle source and destination distributions (to obtain a fair comparison between all the analysed cities we have not consider the Telecom traffic data for Milan here). Thus, for each city, we consider homogeneously distributed source and destination locations and the required road traffic to obtain an order parameter $\eta$ compatible with the congestion effects recorded by INRIX sensing of real traffic. By relating the INRIX value and $\eta$, we are assuming that there exists a relation between the fraction of global congestion and the fraction of extra time wasted reported by INRIX. The obtained results are summarized in Table~\ref{tab:INRIX_vs_HOTSPOT}, which shows that the amount of hotspots is correlated with the INRIX value. This shows evidences that the percentage increase in the average travel time to commute between to city locations is related to the number of congestion hotspots and with the excess of vehicles within the city.

\begin{table}
	\centering
	\caption{Comparison between the INRIX (12 Months) traffic index and the number of hotspots estimated by the proposed model for the most congested cities of the world. }
	\label{tab:INRIX_vs_HOTSPOT}
	\begin{threeparttable}
	\begin{tabular}{lcccc}
	City & INRIX\tnote{a} & hotspots & nodes & links\\
	  \hline
	  Milano & 36.2 & 108 & 6924 & 14315 \\
	  London &  32.4 & 93 & 6378 & 14662 \\
	  Los Angeles & 32.2 & 57 & 6799 & 19368 \\
	  Brussels & 30.5 & 50 & 6645 & 15624 \\
	  Antwerpen & 28.6 & 44 & 6530 & 15252 \\
	  San Francisco & 27.9 & 45 & 8854 & 25530 \\
	  Stuttgart & 21.9 & 34 & 8330 & 19946 \\
	  Nottingham & 21.6 & 28 & 7337 & 16723 \\
	  Karlsruhe & 21.3 & 19 & 4257 & 10379 \\
	  \hline
  	\end{tabular}
	\begin{tablenotes}
		\item[a] The INRIX index is the percentage increase in the average travel time of a commute above free-flow conditions during peak hours, e.g.\ an INRIX index of~30 indicates a 40-minute free-flow trip will take 52~minutes. Each city has been mapped to a graph with the indicated numbers of nodes and links.
        \end{tablenotes}
	\end{threeparttable}
\end{table}

\section{Discussion}

The previous results show that the MCM (Microscopic Congestion Model) can be used to predict the local congestion before and beyond the onset of congestion of a transportation network. Up to the knowledge of the authors, this is the first analytical model that is able to give predictions beyond the onset of congestion where the system is highly non-linear, large fluctuation exists and the amount of vehicles on transit diverge with respect to time. Our model is based on assuming that the growth of vehicles observed in each congested node of the networks is constant, which allowed us to derive a set of balance equations that can accurately predict microscopic, mesoscopic and microscopic variables of the transportation network.

Traffic congestion is a common and open problem whose negative impacts range from wasted time and unpredictable travel delays to a waste of energy and an uncontrolled increase of air pollution. A first step towards the understanding and fight of congestion and its related consequences is the analytical modelling of the congestion phenomena. Here, we have shown that the MCM model is detailed enough to give real predictions considering real traffic data and topology. These results pave the way to a new generation of stilyzed physical models of traffic on networks in the congestion regime, that could be very valuable to assess and test new traffic policies on urban areas in a computer simulated scenario.

\section{Materials and Methods}

\subsection{Microscopic Congestion Model}\label{sec:varsOfModel}

Let node $i$ denote a road junction, edge $a_{ij}$ the road segment between junctions $i$ and $j$, $\nin{i}$ and $\nout{i}$ the sets of ingoing and outgoing neighbouring junctions of junction $i$ respectively, and $S$ the number of junctions in the road network of the city. Incoming vehicles to node $i$ at each time step can be of two types: those coming from other junctions $\nin{i}$ and those that start its trip with node $i$ as its first crossed junction. We consider this second type of incoming vehicles as generated in node $i$. Our Microscopic Congestion Model (MCM) describes the increment of the vehicles per unit time at every junction $i$ of the city, $\Delta{q}_{i}$, as:
\begin{equation}\label{oneLayerMMC}
	\Delta{q}_{i} = g_{i} + \sigma_{i} - d_{i}\,,
\end{equation}
where $g_{i}$ is the average number of vehicles generated in node $i$, $\sigma_{i}$ is the average number of vehicles that arrive to node $i$ from $\nin{i}$ junctions, and $d_{i} \in [0,\tau_i]$ corresponds to the average number of vehicles that actually finish in it, or traverse this junction towards neighboring nodes in $\nout{i}$. Parameter $\tau_i$ represents the maximum routing rate of junction $i$. As described in the main text, we decompose the incoming flux of vehicles $\sigma_i$ to node $i$ as
\begin{equation}\label{sigmadef}
	\sigma_{i} = \sum_{j \in \nin{i}} P_{ji} p_{j} d_{j}\,,
\end{equation}
where $p_i$ is the probability that a vehicle waiting in node $i$ has not arrived to its destination (i.e., it is going to visit at least another junction in the next step) and $P_{ji}$ is the probability that a vehicle crossing node $j$ goes to node $i\in \nout{j}$ in its next step.

Since vehicles just generated in a certain node are not affected by the congestion in the rest of the network, we separate their contributions in the computation of probabilities $p$ and $P$. Thus, we decompose $p_i$ as
\begin{eqnarray}\label{smallP}
	p_{i} & = & p^\expM{gen}_{i}p^\expM{loc}_{i} + (1-p^\expM{gen}_{i}) p^\expM{ext}_{i}\,,
\end{eqnarray}
where the first term accounts for vehicles generated in node $i$ ($p^\expM{gen}_{i}$) whose destination is not $i$ ($p^\expM{loc}_{i}$) and the second term accounts for vehicles not generated in $i$ whose destination is not $i$ ($p^\expM{ext}_{i}$). Supposing trips consist in traveling through two or more junctions we have that $p^\expM{loc}_{i}=1$. Probability $p^\expM{gen}_{i}$ is equal to the fraction of vehicles generated in $i$ with respect to the total amount of incoming vehicles:
\begin{eqnarray}
	p^\expM{gen}_{i} = \frac{g_i}{g_i+\sigma_i}\,.
\end{eqnarray}
Considering the distribution of origins, destinations, the routing strategy and the congestion in the network, probability $p^\expM{ext}_{i}$ can be expressed in terms of the effective node betweenness $\tilde{B}_i$ and the effective vehicle arrivals $\tilde{e}_i$ (the amount of vehicles with destination node $i$ that arrive to node $i$ at each time step):
\begin{eqnarray}
	p^\expM{ext}_{i}=\frac{\tilde{B}_i}{\tilde{B}_i + \tilde{e}_i} \,.
\end{eqnarray}
The effective betweenness $\tilde{B}_i$ of a node $i$ accounts for the expected amount of vehicles each node $i$ receives per unit time considering the routing algorithm and the overall congestion of the network. See {\em Materials and Methods} subsection {\em Effective betweenness in congested transportation networks} for an extended description and computation of the effective node betweenness $\tilde{B}_i$ and the effective vehicle arrivals $\tilde{e}_i$.

In the same spirit, we decompose the probability $P_{ji}$ that a vehicle waiting in node $j$ goes to node $i$ as:
\begin{eqnarray}\label{BigP}
	P_{ji} &=& p^\expM{rgen}_{j}P^\expM{loc}_{ji} + (1-p^\expM{rgen}_{j})P^\expM{ext}_{ji}\,.
\end{eqnarray}
The first term corresponds to the routed vehicles generated in node $j$ ($p^\expM{rgen}_{j}$) that go to node $i$ ($P^\expM{loc}_{ji}$) and the second term to the routed vehicles not generated in $j$ that go to node $i$ ($P^\expM{ext}_{ji}$). Similarly as before, $p^\expM{rgen}_{j}$ can be expressed as the rate between the vehicles generated in $j$ and the total amount of routed vehicles:
\begin{eqnarray}
	p^\expM{rgen}_{j} = \frac{g_j}{g_j+\sigma_j p^\expM{ext}_{j}}\,,
\end{eqnarray}
and, $P^\expM{loc}_{ji}$ and $P^\expM{ext}_{ji}$ can be computed in terms of the normalized effective edge betweenness of the network:
\begin{eqnarray}
	P^\expM{loc}_{ji} &=& \frac{\tilde{E}^\expM{loc}_{ji}}{\displaystyle\sum_{k=1}^S \tilde{E}^\expM{loc}_{jk}} \,, \\
	P^\expM{ext}_{ji} &=& \frac{\tilde{E}^\expM{ext}_{ji}}{\displaystyle\sum_{k=1}^S \tilde{E}^\expM{ext}_{jk}} \,, \label{oneLayerMMC_lastEq}
\end{eqnarray}
where the computation of $\tilde{E}^\expM{loc}_{ji}$ only considers paths that start on node $j$ and $\tilde{E}^\expM{ext}_{ji}$ only considers paths that do not start on node $j$. Equivalently to the effective node betweenness $\tilde{B}_i$, computation of $\tilde{E}^\expM{loc}_{ji}$ and $\tilde{E}^\expM{ext}_{ji}$ consider, if required, all congested junctions in the network, as described in a later section, as well as the distribution of the vehicle sources and destinations. Note that the sum of $E^\expM{loc}_{ji}$ and $E^\expM{ext}_{ji}$ corresponds to the classical edge betweenness. Moreover, $P_{ji}$ is an exact expression before and after the onset of congestion.

Eventually, the MCM is composed by a set of $S$ equations ($\Delta{q}_{i} = g_{i} + \sigma_{i} - d_{i}$), one for each junction, and, in principle, a set of $2S$ unknowns, $\Delta{q}_{i}$ and $d_i$ for each junction. To see that the system is indeed determined we need to note that for congested junctions $\Delta q_{i} > 0$ and, thus, after the transient state, $d_{i}=\tau_i$. For the non-congested junctions we have that $\Delta q_{i}=0$ and consequently $d_i =  g_i + \sigma_i$. In conclusion, for any node $i$, either $d_i=\tau_i$ or $d_i=g_i + \sigma_i$ which reduces the amount of unknowns to $S$.

To solve the model given a fixed generation rate $g_i$, we start by considering that no junction is congested and we solve the set of equations Eqs.~\ref{oneLayerMMC}--\ref{oneLayerMMC_lastEq} by iteration. It is possible that some nodes exceed their maximum routing rate. If this is the case, we set the node with maximum $d_i$ as congested and we solve the system again. This process is repeated until no new junction exceeds its maximum routing rate.

\subsection{Onset of congestion using the Microscopic Congestion Model}

Most of the works that consider static routing strategies assume that the generation rate of vehicles is the same for all nodes, $g_i=\rho$. In that case, it is possible to compute the critical generation rate $\rho_c$ such that for any generation rate $\rho > \rho_c$ the network will not be able to route or absorb all the traffic\cite{Guimera2002OptimalTopologies,Arenas01PRL,Martino09PRE,Chen12MPE,Arenas03LNP,Arenas2010OptimalInfoTransm}. After this point is reached, the amount of vehicles $Q(t)$ in the network will grow proportionally with time, $Q(t) \propto t$, since some of the vehicles get stacked in the queues of the nodes. This transition to the congested state is characterized using the following order parameter:
\begin{equation}\label{oneLayerOrderParameter}
	\eta(\rho)= \lim_{t\rightarrow\infty}\frac{\langle{\Delta Q}\rangle}{\rho S}\,,
\end{equation}
where $\langle\Delta Q\rangle$ represents the average increment of vehicles per unit of time in the stationary state. Basically, the order parameter measures the ratio between in-transit and generated vehicles.

In the non-congested phase, the amount of incoming and outgoing vehicles for each node can be computed in terms of the node's algorithmic betweenness $B_i$, see ref.~\cite{Guimera2002OptimalTopologies}. In particular,
\begin{equation}\label{oneLayerSigma}
	\sigma_{i} = \rho\(\frac{B_i}{S-1} + 1\)\,,
\end{equation}
where the second term inside the parentheses accounts for the fact that, in our model, vehicles are also queued at the destination node, unlike in ref.~\cite{Guimera2002OptimalTopologies}. When no junction is congested we have that $\Delta{q}_{i} = 0$ for all nodes and consequently
\begin{equation}\label{oneLayerD^i}
	d_i = \rho + \sigma_i = \rho\(\frac{B_i}{S-1} + 2\)\,.
\end{equation}
A node $i$ becomes congested when it is required to process more vehicles than its maximum processing rate, $d_i > \tau$. Thus, the critical generation rate at which the first node, and so the system, reaches congestion is:
\begin{equation}\label{rhocFirstNode}
	\rho_{c} = \min_i \frac{\tau\(S-1\)}{B_i + 2(S-1)}\,.
\end{equation}

This is one of the most important analytical results on transportation networks with static routing strategies. In the following, we show that we can recover Eq.~(\ref{oneLayerSigma}) before the onset of congestion using our MCM approach. After substitution of the expression of the probabilities in Eq.~\ref{sigmadef}:
\begin{equation}
	\sigma_i = \sum\limits_j \frac{\rho(B_j + S-1) P^\expM{loc}_{ji} + \sigma_j B_j P^\expM{ext}_{ji}}{(\rho+\sigma_j)(B_j + S-1)} d_j \,,
\end{equation}
and, given we do not have congestion (i.e., $d_j = \rho + \sigma_j$), it simplifies to
\begin{equation}\label{eq:sigmaMCM}
	\sigma_i = \sum\limits_j \frac{\rho(B_j + S-1) P^\expM{loc}_{ji} + \sigma_j B_j P^\expM{ext}_{ji}}{B_j + S-1}\,.
\end{equation}
Equation (\ref{eq:sigmaMCM}) in matrix form becomes
\begin{equation}
	(I - M)\boldsymbol{\sigma} = \rho\boldsymbol{\pi} \,,
\end{equation}
where
\begin{eqnarray}
	M_{ij} &=& \frac{B_j P^\expM{ext}_{ji}}{B_j + S - 1}\,,\\
	\pi_i &=& \sum_j P^\expM{loc}_{ji}\,,
\end{eqnarray}
and then
\begin{equation}
	\boldsymbol{\sigma} = \rho(I-M)^{-1}\boldsymbol{\pi}\,.
\end{equation}
This expression can be shown to be equivalent to Eq.~(\ref{oneLayerSigma}) by using the following relationship between node and edge betweenness:
\begin{equation} \label{nodeEdgeBWRelation}
	B_i + (S-1)=  \sum_{j} \left(B_{j} P^\expM{ext}_{ji} + (S-1)P^\expM{loc}_{ji}\right)\,.
\end{equation}
The right hand side corresponds to the accumulated fractions of paths that pass through the neighbors of node $i$ and then go to $i$. Each neighbor contributes with two terms, the paths that go through $j$ coming from other nodes, and the paths that start in $j$.

\subsection{Effective betweenness in congested transportation networks}\label{sec:effectiveBW}

The effective betweenness $\tilde{B}_i$ of a node $i$, as defined in ref.~\cite{Guimera2002OptimalTopologies}, accounts for the expected amount of vehicles each node $i$ receives per unit time. When the network is not congested and the vehicle generation rate $g_i$ is equal for all nodes, $g_i = \rho$, the number of vehicles each node receives can be obtained using Eq.~\ref{oneLayerSigma}. However, if the network is congested, the traffic dynamics becomes highly non-linear and the value of $\sigma_i$ computed in Eq.~\ref{oneLayerSigma} becomes a poor approximation.

Suppose we focus on a particular congested node $j^{\ast}$ of the network. For $j^{\ast}$, being congested means that it is receiving more vehicles that the ones it can process and route. In particular, from the $\sigma_{j^{\ast}}+g_{j^{\ast}}$ vehicles that arrive to the node, only $\tau_{j^{\ast}}$ can be processed at each time step.

Therefore, the contribution to the effective betweenness $\tilde{B}_i$ of the paths from a source/destination pair, $(s,t)$, that traverse the congested node $j^{\ast}$ before reaching $i$, must be rescaled by the fraction of processable vehicles:
\begin{equation}\label{rescalingEffBetweenness}
	s_{j^{\ast}} = \frac{\tau_{j^{\ast}}}{\sigma_{j^{\ast}} + g_{j^{\ast}}}\,.
\end{equation}
When a path traverses multiple congested nodes ${j^{\ast}},{k^{\ast}},\dots$, the remaining fraction of paths that will reach the target node will be the result of the application of the multiple re-scalings $s_{x^{\ast}}$.

The computation of $s_{j^{\ast}}$ is not straightforward. In general, $\sigma_i$ is not known after the onset of congestion and depends on the effective betweenness that requires, at the same time, to know the $s_{j^{\ast}}$ fraction for all congested nodes. Thus, an iterative calculation is needed to fit all the parameters at the same time as we do in our Microscopic Congestion Model.

The effective arrivals $\tilde{e}_i$ account for the amount of vehicles with destination node $i$ that arrive to node $i$ at each time step. This value in the non-congested phase can be obtained, considering homogeneous source and destination nodes, as
\begin{equation}\label{effectiveArrivals}
	e_i = \rho(S-1)\,.
\end{equation}
However, congestion affects the variable $e_i$ as well, and needs to be corrected accordingly using the same procedure presented above.

\subsection{Traffic Dynamics} \label{sec:traffic_dynamics}

To simulate the traffic dynamics of the road network, we assign a first-in-first-out queue to each junction that simulates the blocking time of vehicles before they are allowed to cross it and continue their trip. We suppose these queues have infinite capacity and a maximum processing rate that simulates the physical constraints of the junction. Vehicles origins and destinations may follow any desired distribution. In this work, we have considered two distributions: a random uniform distribution for the synthetic experiments, and one obtained considering the ingoing and outgoing flux of vehicles of the city of Milan.  At each time step (of 1 minute duration) vehicles are generated and arrive to their first junction. During the following time steps, vehicles navigate towards their destination following any routing strategy. Here, we have used two different routing strategies: shortest-path and random local search.

For the particular case of simulating traffic in the city of Milan we assume a traffic dynamics similar to the ``Home-to-Work'' travel pattern where vehicles arrive from the outskirts of the city, go to the city center and then return to the outskirts. Specifically, in our simulation, traffic is generated in the peripheral junctions of the network, goes to a randomly selected junction within the city and then returns back to a randomly selected peripheral junction. We do not consider trips with origin and destination inside the city center since public transportation systems (e.g., train or subway) usually constitute a better alternative than private vehicles for those trips.

The maximum crossing rate of each junction $\tau_i$ accounts, among others, for the existence of traffic lights governing the junction, the width of the street as well as its traffic. We have not been able to get this information for the studied cities, and consequently we cannot set to each junction its precise value. Instead, without loss of generality and for the sake of simplicity, we set to all junctions the same maximum crossing rate, $\tau_i = 15$ (an estimation of the average of their real values).

\section{Acknowledgements}
This work has been supported by Ministerio de Econom\'{\i}a y Competitividad (Grant FIS2015-71582-C2-1) and European Comission FET-Proactive Projects MULTIPLEX (Grant 317532). A.A.~also acknowledges partial financial support from the ICREA Academia and the James S. McDonnell Foundation.


\end{document}